\documentclass[aps,prl,twocolumn,superscriptaddress]{revtex4-2}

\usepackage{empheq}
\usepackage{graphicx}
\usepackage{epsfig}
\usepackage{amsmath,amsthm}
\usepackage{color}
\usepackage{verbatim}
\usepackage{ulem}
\usepackage{siunitx}
\usepackage{lineno}

\begin{document}

\preprint{unpublished manuscript}


\title{Identifying Switching of Antiferromagnets by Spin-Orbit Torques}

\author{Martin Jourdan}
			\email{Jourdan@uni-mainz.de}
	\affiliation{Institut f\"ur Physik, Johannes Gutenberg-Universit\"at, Staudinger Weg 7, 55128 Mainz, Germany}
\author{Jonathan Bl\"a{\ss}er}
	\affiliation{Institut f\"ur Physik, Johannes Gutenberg-Universit\"at, Staudinger Weg 7, 55128 Mainz, Germany}
\author{Guzm\'an Orero G\'amez}
	\affiliation{Institut f\"ur Physik, Johannes Gutenberg-Universit\"at, Staudinger Weg 7, 55128 Mainz, Germany}
\author{Sonka Reimers}
	\affiliation{Institut f\"ur Physik, Johannes Gutenberg-Universit\"at, Staudinger Weg 7, 55128 Mainz, Germany}
\author{Lukas Odenbreit}
	\affiliation{Institut f\"ur Physik, Johannes Gutenberg-Universit\"at, Staudinger Weg 7, 55128 Mainz, Germany}
\author{Miriam Fischer}
	\affiliation{Institut f\"ur Physik, Johannes Gutenberg-Universit\"at, Staudinger Weg 7, 55128 Mainz, Germany}
\author{Yuran\,R. Niu}
	\affiliation{MAX IV Laboratory, Fotongatan 8, 22484 Lund, Sweden}
\author{Evangelos\,Golias}
	\affiliation{MAX IV Laboratory, Fotongatan 8, 22484 Lund, Sweden}
\author{Francesco Maccherozzi}
	\affiliation{Diamond Light Source, Chilton OX11 0DE, United Kingdom}
\author{Armin Kleibert}
	\affiliation{Paul Scherrer Institut, CH-5232 Villigen PSI, Switzerland}
\author{Hermann Stoll}
	\affiliation{Institut f\"ur Physik, Johannes Gutenberg-Universit\"at, Staudinger Weg 7, 55128 Mainz, Germany}
	\affiliation{Max Planck Institute for Intelligent Systems, Heisenbergstr. 3, 70569 Stuttgart, Germany}
\author{Mathias Kl\"aui}
	\affiliation{Institut f\"ur Physik, Johannes Gutenberg-Universit\"at, Staudinger Weg 7, 55128 Mainz, Germany}

\begin{abstract}
Antiferromagnets are promising candidates for ultrafast spintronic applications, leveraging current-induced spin-orbit torques. However, experimentally distinguishing between different switching mechanisms of the staggered magnetization (N{\'e}el vector) driven by current pulses remains a challenge. In an exemplary study of the collinear antiferromagnetic compound Mn$_2$Au, we demonstrate that slower thermomagnetoelastic effects predominantly govern switching over a wide parameter range. In the regime of short current pulses in the nanosecond range, however, we observe fully N\'eel spin-orbit torque driven switching. We show that this ultrafast mechanism enables the complete directional alignment of the N\'eel vector by current pulses in device structures.     
\end{abstract}



\maketitle

* email: Jourdan@uni-mainz.de\\

\section{Introduction}
Antiferromagnets (AFMs) have been proposed as a groundbreaking platform in spintronics, enabling information storage through the encoding of data in the alignment of the staggered magnetization, also termed N\'eel vector \cite{Mac11,Bal18,Jun18,Jung18}. The integration of antiferromagnets (AFMs) as active components in spintronic devices can offer significant advantages, such as stability against external magnetic fields requiring additionally magnetic anisotropy \cite{Mac17}. In particular, the magnetic domain configuration of our samples is stable in magnetic fields up to 30~T \cite{Sap18}. Furthermore, the intrinsically fast terahertz (THz) dynamics of AFMs \cite{Rez19} enables the manipulation of magnetic order by THz pulses \cite{Kam11}. 

Most approaches to the information-writing process, specifically the realignment of the N{\'e}el vector, focus on leveraging current pulse-induced spin-orbit torques (SOTs), as this mechanism is anticipated to enable a potentially ultrafast operation. Rotation of the N{\'e}el vector is envisioned to be driven by a strong exchange torque, generated when the AFM sublattices are tilted out of their originally antiparallel alignment \cite{Bal18}. In principle, such spin-orbit torques (SOTs) can arise from interfaces between antiferromagnets (AFMs) and non-magnetic heavy metals, as well as from the bulk of metallic AFMs with inversion symmetry breaking in their combined crystallographic and magnetic structure, a mechanism referred to as N{\'e}el spin-orbit torque (NSOT) \cite{Zel14, Sal19, Sel22}).

However, experimental evidence for the contribution of NSOTs to current-induced N{\'e}el vector reorientation remains limited. These indications primarily consists of observations of current-polarity-dependent motion in a small fraction of domain walls in CuMnAs \cite{Wad18,Ami23} and Mn$_2$Au \cite{Rei23}, and in a complex directional dependence of small domain modifications in NiO/Pt \cite{Sch24}.
 
On the other hand, an alternative effect has been identified that also induces N{\'e}el vector reorientation in antiferromagnets, albeit at a much slower rate compared to SOT-driven processes. This thermomagnetoelastic mechanism arises from current-induced heating, which generates strain and modifies the magnetic anisotropy, ultimately leading to N{\'e}el vector reorientation \cite{Bal20, Mee21}. Distinguishing between these two mechanisms is challenging, yet it is crucial to experimentally verify that current pulse-driven SOT can serve as an ultrafast switching mechanism in antiferromagnetic spintronics.

In principle, there is also a third mechanism to be considered, which is based on inhomogeneous current heating due to the anisotropic magnetoresistance (AMR) of AFM domains with different N{\'e}el vector orientation. Within this model, the domains with higher resistance, i.\,e.\,temperature, should be removed \cite{Sel16}. However, for the compound investigate here, this mechanism can be excluded, because the AMR of Mn$_2$Au results in a higher resistance for a perpendicular alignment of N{\'e}el vector and current \cite{Bod20}, which is the final configuration after the switching discussed below.  

In this study, we focus on tetragonal Mn$_2$Au \cite{Bar13}, which shows collinear AFM ordering with four equivalent easy $\langle 110 \rangle$-directions. An NSOT was predicted for this exemplary compound \cite{Zel14}, however, it was also shown that its AFM domain configuration also be modified by externally applied strain \cite{Che19, Gri22}. 

We investigate current pulse driven N{\'e}el vector switching in Mn$_2$Au and demonstrate that for pulses of 10~µs or longer, the reorientation process is predominantly driven by the thermomagnetoelastic effect. In contrast, for pulses shorter than 100~ns, we observe switching of AFM domains driven purely by NSOT. Our conclusions are drawn from the analysis of the N{\'e}el vector orientation relative to the current direction, which, for a given geometry, exhibits distinct characteristics for the two mechanisms. Additionally, we demonstrate that an effective NSOT results the alignment of the N{\'e}el vector along a specific direction, in contrast to the bidirectional alignment observed with a strain-induced easy axis.

\section{Results and discussion}

{\bf Direction of NSOT and thermomagnetoelastic N\'eel vector reorientation}

In this section we will show how the choice of the sample geometry selects either a cooperation or a competition of the thermomagnetoelastic and of the NSOT switching mechanisms.

We consider a thin film geometry with a magnetic hard axis perpendicular to the sample plane.
The direction of NSOT driven N{\'e}el vector reorientation is fully determined by the direction of the driving current ${\mathbf J}$. Following the derivation by {\v Z}elezn{\'y} et al. \cite{Zel14}, the staggered effective fields $\mathbf{B^\mathrm{A}}=-\mathbf{B^\mathrm{B}}$ at the AFM sublattices A and B are defined by the cross product of the normal ${\mathbf z}$ on the easy thin film plane and the direction of the in-plane current ${\mathbf J}$ (see Fig.\,1):
\begin{equation}\label{staggered B}
\begin{split}
\mathbf{B^\mathrm{A}} \propto + \mathbf{z} \times \mathbf{J} \\\
\mathbf{B^\mathrm{B}} \propto - \mathbf{z} \times \mathbf{J}
\end{split}
\end{equation}
With this staggered effective field, the field like torques $\mathbf{T^\mathrm{A}}$ and $\mathbf{T^\mathrm{B}}$ act for both AFM sublattices with antiparallel magnetizations $\mathbf{M^\mathrm{A}}$ = $-\mathbf{M^\mathrm{B}}$ in the same direction, i.\,e.\,$\mathbf{T^\mathrm{A}}=\mathbf{T^\mathrm{B}}$:
\begin{equation}\label{staggered B}
\begin{split}
\mathbf{T^\mathrm{A}} \propto \mathbf{M^\mathrm{A}} \times (+ \mathbf{z} \times \mathbf{J}) \\\
\mathbf{T^\mathrm{B}} \propto \mathbf{-M^\mathrm{A}} \times (- \mathbf{z} \times \mathbf{J})
\end{split}
\end{equation}
The resulting canting of the sublattice magnetizations in the same direction out of the easy-plane generates an exchange torque $\mathbf T_{\rm E}$ driven rotation of the N{\'e}el vector around ${\mathbf z}$, whose direction depends on the canting direction \cite{Bal18}. $\mathbf{T^\mathrm{A/B}}$ is maximum for antiparallel/ parallel alignment and zero for perpendicular alignment of the N\'eel vector to the current direction. If the magnetization directions of the AFM sublattices are swapped ($\mathbf{M^\mathrm{A/B}} \rightarrow -\mathbf{M^\mathrm{A/B}}$), corresponding to a sign change of the N\'eel vector, the torques change sign as well as shown in Eq.\,2. Thus, as visualized in Fig.\,1, NSOT driven by a current parallel to the magnetic moments will reorient the N{\'e}el vector of $180^{\rm o}$-domains in the same direction perpendicular to the current.
\begin{figure}
\includegraphics[width=0.7\columnwidth]{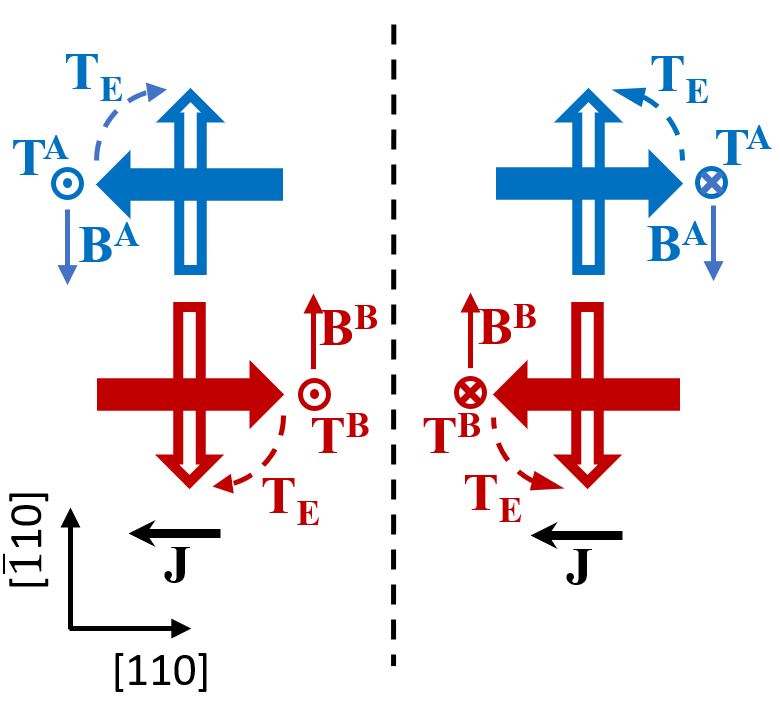} 
\caption{\label{Fig1} 
{\bf Schematic representation of NSOT driven N{\'e}el vector switching.} The initial configuration shows two antiparallel orientations of N{\'e}el vectors ($180^{\rm o}$-domains) corresponding to swapped sublattice magnetizations $\mathbf{M_{A/B}}$ (filled arrows). The sublattice sites are connected by a crystallographic inversion. An NSOT driven by the current $\mathbf{J}$ reorients the N{\'e}el vectors $\mathbf{N}=\frac{1}{2} (\mathbf{M_A}-\mathbf{M_B})$ of both domains in the same direction indicated by the open arrows, thereby removing the $180^{\rm o}$-domain wall indicated by the dashed line.}
\end{figure}

Tetragonal Mn$_2$Au is characterized by a strong magnetic anisotropy with a hard c-axis. The magnetic anisotropy within the (001)-plane, which is much smaller, favors two mutually orthogonal magnetic easy axes along the $\langle 110 \rangle$ crystallographic directions. This results in a domain configuration of as grown epitaxial Mn$_2$Au(001) thin film of sub-$\mu$m$^2$ sized domains with four $\langle 110 \rangle$ directions of the N{\'e}el vector \cite{Bom21}. Based on the discussion above, an NSOT arising from a current along the [$\bar{1}\bar{1}$0] direction is expected to switch all domains with N{\'e}el vectors aligned parallel or antiparallel to this direction, not only perpendicular to the current but also into the same direction, such as the [$\bar{1}$10] direction (Fig.\,1). As no NSOT acts on the domains with N{\'e}el vectors aligned perpendicular to the current direction, those with N{\'e}el vectors pointing in the antiparallel [1$\bar{1}$0] direction are unaffected, and $180^{\rm o}$ domain walls will persist. Thus, after a single current pulse through an ensemble of domains representing all four easy $\langle110\rangle$-directions, the N{\'e}el vector of all, but those with [1$\bar{1}$0]-orientation, will point in the [$\bar{1}$10] direction perpendicular to the current. However, these remaining domain walls are expected to be eliminated by a second current pulse now applied along the [1$\bar{1}$0] direction (which is perpendicular to the direction of the first pulse), as the NSOT will again rotate the antiparallel N{\'e}el vectors in opposite directions around the thin film normal by 90$^{\rm o}$. Thus, NSOT is characterized by the complete alignment of the N{\'e}el vector, resulting in the formation of one large AFM domain within the active area of a switching device.
 
In contrast, the thermomagnetoelastic mechanism results in the formation of one axis, along which the N{\'e}el vector aligns both parallel and antiparallel, i.\,e.\,it favors the formation of $180^{\rm o}$ domains. Thermomagnetoelastic switching arises from current-induced heating and the resulting anisotropic thermal expansion of patterned thin-film samples clamped to thick insulating substrates. This can be illustrated by considering a thin, patterned metallic stripe heated by a current pulse. The normal thermal expansion of the stripe, associated with its temperature increase, is constrained by its clamping to the substrate, resulting in strain. However, this constraint is stronger in the direction parallel to the stripe than in the perpendicular direction, leading to compressive strain along the stripe. Although the magnitude and sign of the magnetoelastic energy of Mn$_2$Au are unknown, it is clear that the magnetic anisotropy is modified, driving a reorientation of the N{\'e}el vector. The axis of thermomagnetoelastic N{\'e}el vector alignment depends on the specific device geometry, as demonstrate below through simulations. Consequently, it can be either parallel or perpendicular to the current direction, depending on the sample geometry, which serves as a clear criterion for distinguishing it from NSOT-driven N{\'e}el vector reorientation.

The direction of the strain resulting from a current pulse along a specific crystallographic direction can be varied by selecting an appropriate sample geometry. This means that the axis along which the thermomagnetoelastic effect aligns the N\'eel vector is adjustable, in contrast to the case of the NSOT mechanism, which always aligns the N\'eel vector perpendicular to the current direction.
\begin{figure*}
\includegraphics[width=1.5\columnwidth]{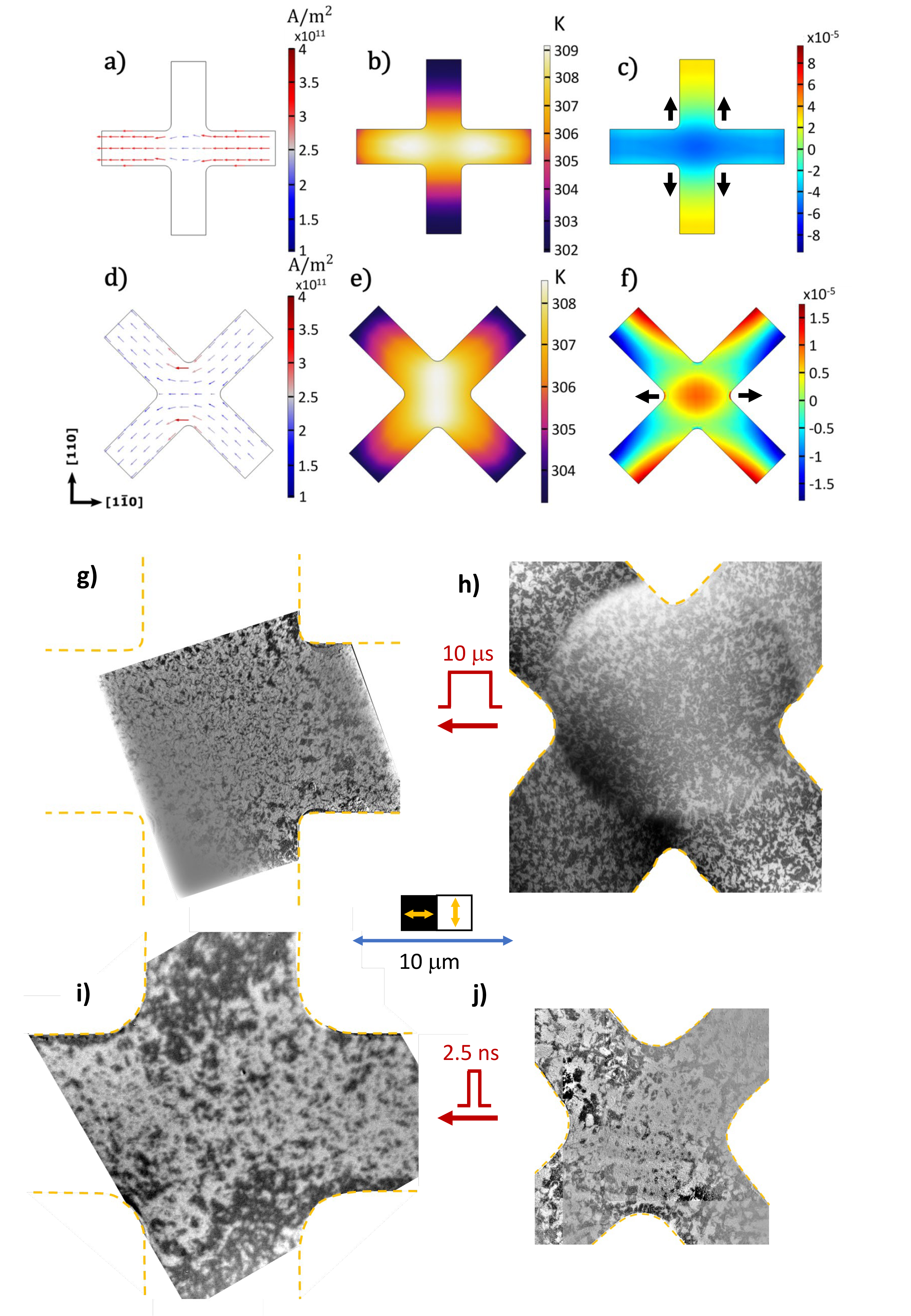} 
\caption{\label{Fig2} 
{\bf Simulations and XMLD-PEEM images of current induced N{\'e}el vector reorientation.} The top row (panels {\bf a}–{\bf c}) presents COMSOL \cite{COM} simulations for a $0^{\rm o}$ cross configuration. In panel {\bf a}, the current flows from the right arm to the left arm, panel {\bf b} shows the associated temperature distribution after a 10~$\mu$s pulse, and panel {\bf c} depicts the resulting strain (black arrows indicate the direction of larger expansion), represented as the difference in relative expansion between the $[1\bar{1}0]$ and $[110]$ directions (see color-bar). Panels {\bf d}–{\bf f} display the corresponding simulations for a cross rotated by $45^{\rm o}$ relative to the easy directions, where the current flows from both right arms to both left arms. Panels {\bf g} and {\bf h} (composed of 5 images) provide XMLD-PEEM images for experimental comparison of the two geometries after a 10~$\mu$s pulse ({\bf g}: $3.3\times10^{11}$~A/cm$^2$, {\bf h}: $3.7\times10^{11}$~A/cm$^2$), while panels {\bf i} and {\bf j} show the corresponding images following a $2.5$~ns pulse ({\bf i} and {\bf j}: $2.5\times10^{12}$~A/cm$^2$). The dimensions of the cross shown in panel {\bf j} are reduced to enable larger current densities. Due to reduced experimental resolution, the $180^{\rm o}$-domain walls appear blurred in panel {\bf i}}. 
\end{figure*}
The typical geometry for current induced N\'eel vector switching is cross-shaped. As shown in Fig.\,2, this cross can be oriented with its arms parallel to the easy $\langle 110 \rangle$ axes ($0^{\rm o}$ cross) or parallel to the in-plane hard $\langle 100 \rangle$ axes ($45^{\rm o}$ cross). In both cases, the current direction in the center of the cross is selected to be parallel to the $[1\bar{1}0]$ direction as shown by the COMSOL \cite{COM} simulations in panels {\bf a} and {\bf d} of Fig.\,2. The simulated spatially dependent temperature distribution at the end of a current pulse with a width of 10~$\mu s$ is shown in panels {\bf b} and {\bf e}. The corresponding spatial distributions of the strain, i.e.\,the relative difference in expansion between the $[1\bar{1}0]$ and the $[110]$ directions are shown in panels {\bf c} and {\bf f}. Please note that the sign of the strain is different for the two geometries, i.e.\,the centers of the crosses are strained (elongated) for the $0^{\rm o}$ cross perpendicular and for the $45^{\rm o}$ cross parallel to the current direction. This is related to the rotated orientation of the heated zones in the two geometries. Furthermore, the absolute value of the current heating induced strain is about a factor of five smaller for the $45^{\rm o}$ cross geometry. The magnitude of the strain generally decreases with decreasing pulse widths and current density. 

The materials parameters required for the simulations (heat capacity, thermal conductivity, thermal expansion coefficient, Young’s modulus, and Poisson ratio) are unknown for Mn$_2$Au. Thus, typical values for metals were chosen. For details, see Supplemental Material, section IV.
\\

{\bf Imaging the switched domain configuration}

We now present the corresponding experimental results, obtained by investigating patterned epitaxial Mn$_2$Au(001) thin films using X-ray magnetic linear dichroism – photoelectron emission microscopy (XMLD-PEEM, see {\it Methods}). The XMLD signal, measured at the Mn-L$_{2,3}$-edge, ideally produces a black-and-white contrast for the perpendicular $\langle 110 \rangle$-axes along which the N{\'e}el vector is aligned. However, due to a small but unavoidable small energy dispersion of the photon beam over the illuminated sample area, the observed XMLD contrast exhibits spatial variation over the images.

The as-prepared AFM domain configuration of the epitaxial Mn$_2$Au(001) thin films is discussed in detail in Ref.\,\cite{Rei24}; a typical example, featuring an equal population of sub-$\mu$m$^2$ sized domains, each with the N\'eel vector aligned along one of the four easy $\langle 110 \rangle$-directions, is presented in Fig.\,S1 of the Supplemental Material \cite{Supp}.

All switched samples shown here were initially pulsed along one of the easy $\langle 110 \rangle$-directions, followed by pulses along a perpendicular easy direction.

Fig.\,2, panels {\bf g} to {\bf j} display magnetic microscopy images of the AFM domain configuration obtained after applying current pulses with variable length in the configurations shown in panels {\bf a} and {\bf d}. In the XMLD-PEEM setup, horizontal alignment of the N{\'e}el vector appears dark, while vertical alignment appears bright. 

Current pulses with a duration of 10~${\mu}$s, applied along the same crystallographic [1$\bar{1}$0] direction, result in a strong predominance of domains with the N{\'e}el vector aligned perpendicular to the current direction (bright) for the 0$^{\rm o}$ cross configuration shown in panel {\bf g}. In contrast, for the 45$^{\rm o}$ cross configuration shown in panel {\bf h}, the N{\'e}el vector orientation parallel to the current direction (dark) is dominant. This means that the same current pulse direction leads to mutually perpendicular alignment of the N\'eel vector in the center of the 0$^{\rm o}$ and 45$^{\rm o}$ crosses, corresponding to the perpendicular directions of the current-induced strain shown in panels {\bf c} and {\bf f}. Thus, these results provide strong evidence for the thermomagnetoelastic switching mechanism. Moreover, our experiments rule out NSOT as the driver of the reorientation process for pulse widths of 10~$\mu$s or longer (see also Supplemental Material, Fig.\,S2), as the direction of the NSOT depends solely on the current direction.

However, for current pulses with a length of $2.5$~ns, both centers of the 0$^{\rm o}$ and the 45$^{\rm o}$ crosses shown in panels {\bf i} and {\bf j}, display a strong predominance of N{\'e}el vector aligned perpendicular to the current direction by appearing bright in the XMLD-PEEM images. This result rules out the thermomagnetoelastic switching mechanism but is fully consistent with NSOT driven N\'eel vector reorientation.
\begin{figure*}
\includegraphics[width=1.5\columnwidth]{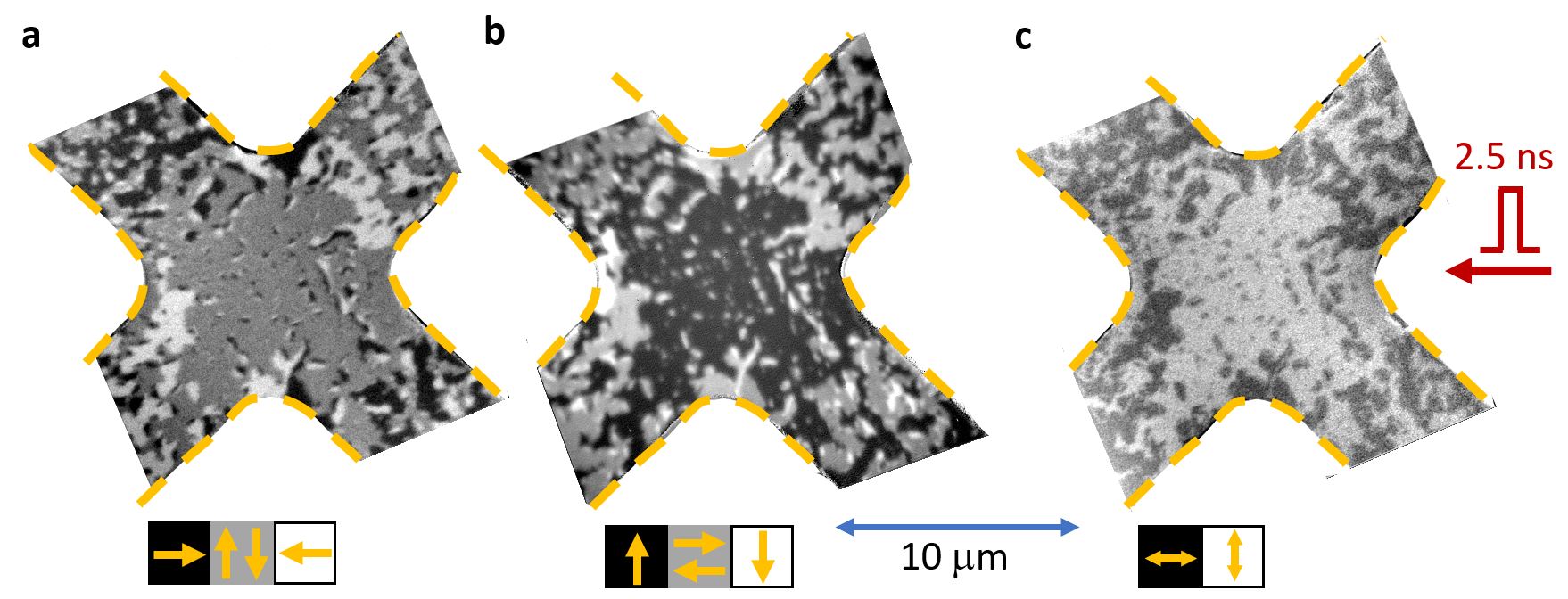} 
\caption{\label{Fig1} 
{\bf XMCD- and XMLD-PEEM images of current induced N{\'e}el vector reorientation of Mn$_2$Au/Ni$_{80}$Fe$_{20}$.} Panels {\bf a} and {\bf b} show the magnetization orientation of the Ni$_{80}$Fe$_{20}$ as obtained through XMCD-PEEM imaging at the Fe-edge, with two different  x-ray incidence directions. Panel {\bf c} represents the corresponding XMLD-PEEM image at the Mn-edge, captured after a 2.5~ns current pulse ($3.3\times10^{12}$~A/cm$^2$), as indicated in the figure.} 
\end{figure*}

Compelling evidence for NSOT-induced switching by nanosecond current pulses is provided by the observation of a predominantly single large domain in the center of the 45$^{\rm o}$ cross, as shown in Fig.\,2{\bf j}. This is because NSOT drives the N{\'e}el vector to align along a specific direction removing 180$^{\rm o}$ domain walls. In contrast, after a 10~$\mu$s pulse, the center of the thermomagnetoelastically switched cross contains multiple closed loops of 180$^{\rm o}$-domain walls with a width close to the resolution limit in the XMLD-PEEM image as shown in Fig.\,2{\bf g} and discussed in more detail in section II of the Supplemental Material.

To directly visualize the alignment of the N{\'e}el vector after the $2.5$~ns pulses, we have additionally investigated 45$^{\rm o}$ crosses of Mn$_2$Au(40~nm)/Ni$_{80}$Fe$_{20}$(4~nm) bilayers. As described in ref.\,\cite{Bom21}, these bilayers behave like an exchange spring and exhibit a strong collinear exchange coupling between the N{\'e}el vector and the magnetization of the ferromagnetic Ni$_{80}$Fe$_{20}$. Consequently, externally driven rotations of either vector induce a coupled rotation of the other. By leveraging this coupling, we indirectly determine the direction of the N{\'e}el vector using XMCD-PEEM at the Fe edge, which reveals the orientation of the Ni$_{80}$Fe$_{20}$ magnetization (Fig.\,3, panels {\bf a} and {\bf b}). XMLD-PEEM at the Mn edge (panel {\bf c}) confirms the assumed one-to-one correspondence between the AFM and FM domains also in the switched state.

These results highlight the unique capability of the NSOT to fully align the N{\'e}el vector and thereby to create a large single AFM domain from an initially multidomain state. 
\\

{\bf Electric characterization of switching}

We employed both XMLD-PEEM imaging and anisotropic magnetoresistance measurements, associated with N\'eel vector reorientation, to identify the switched states (see {\it Methods}). Thermomagnetoelastic N\'eel vector reorientation was observed for current pulse durations of 1~ms (Supplementary Information, Fig.\,S2) and 10~$\mu$s (Fig.\,2{\bf h}). In this regime, current pulses were applied in situ within the PEEM, enabling microscopic observation of changes in the AFM domain structure immediately after each pulse as the current density increased. Pulse widths in the range of 10~$\mu$s to 100~ns were applied outside the PEEM, as the sample stage wiring within the microscope does not support the impedance matching necessary for these shorter pulses. To pre-characterize the switched state, we measured the transverse resistance of the samples prior to PEEM imaging. After each pulse application at progressively increasing current densities, the transverse resistance (planar Hall effect) was recorded. In these measurements, the increasing fraction of the switched area at the center of the cross-shaped devices is reflected as a corresponding increase or decrease in resistance. Once the resistance change saturates, the sample is considered fully switched and can be imaged ex situ using XMLD-PEEM.
\begin{figure*}
\includegraphics[width=1.5\columnwidth]{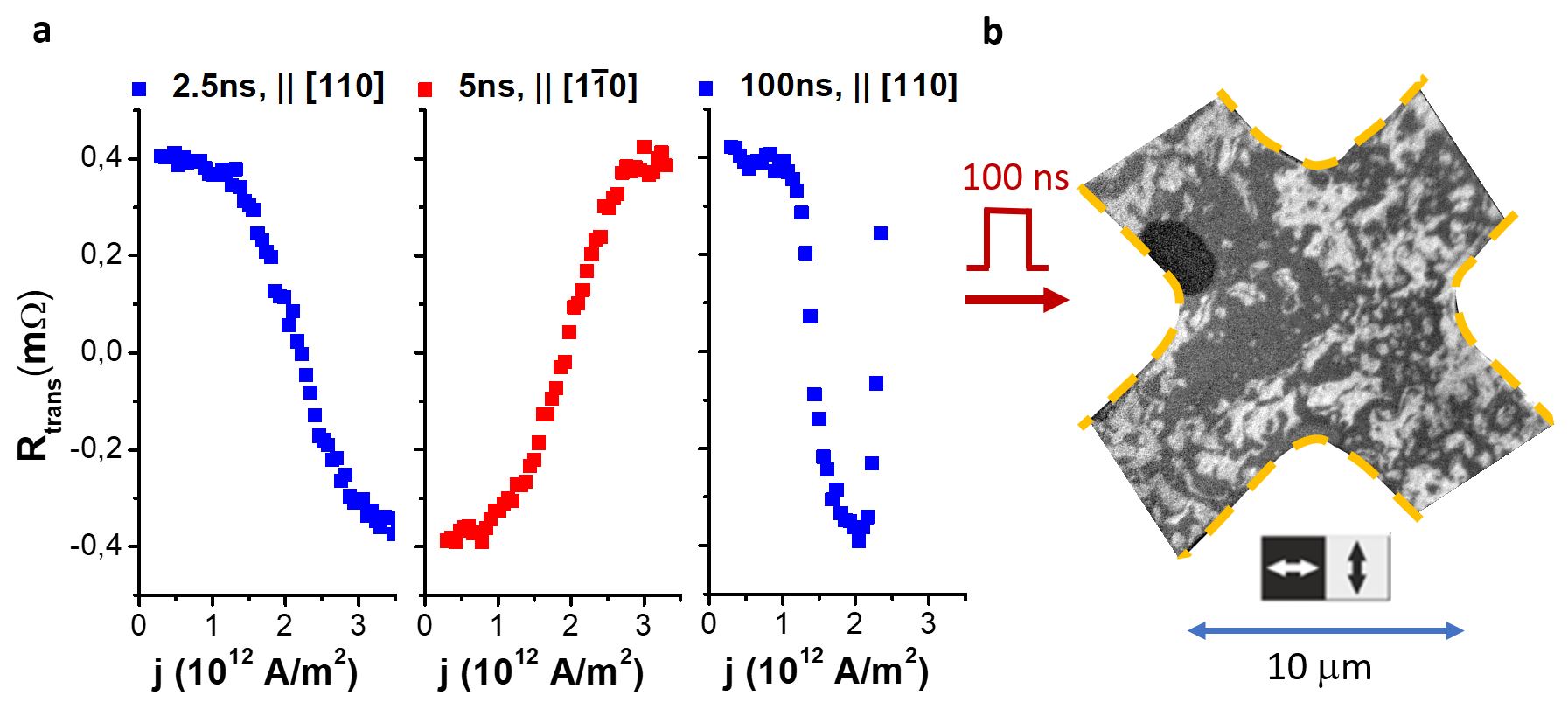} 
\caption{\label{Fig1} 
{\bf Effect of N{\'e}el vector reorientation on the Mn$_2$Au resistance and reversed switching.} Panel {\bf a} shows the transverse resistance of a 45$^{\rm o}$ Mn$_2$Au cross in response to sequences of current pulses along alternating perpendicular easy directions, with the pulse width indicated in the figure. After the final pulse sequence with a width of 100~ns, which resulted in a reversal of the resistance change, the sample was imaged using XMLD-PEEM, as shown in panel {\bf b}.} 
\end{figure*}

Complete NSOT switching was observed at current densities ranging from $2\times 10^{12}$~${\rm A/m^2}$ (100~ns pulses) to $3\times 10^{12}$~${\rm A/m^2}$ (2.5~ns pulses) as shown in Fig.\,4, panel {\bf a}. These values exceed by more than a factor of five the current density of $4\times 10^{11}$~${\rm A/m^2}$ required for thermomagnetoelastic switching with 10~$\mu$s pulses. As the thermally generated strain induced by current pulses decreases with shorter pulse widths, thermomagnetoelastically driven switching eventually requires higher current densities than NSOT switching. Only in this cross-over regime switching driven by both mechanisms can be observed at different current densities: For a pulse width of 100~ns, we observed NSOT switching at lower current densities, followed by thermomagnetoelastic switching at higher current densities, as indicated by the transverse resistance measurements in Fig.\,4{\bf a}. For 100~ns pulse widths, the induced resistance changes initially follow the behavior observed for nanosecond pulses but subsequently reverse sign. In this regime, saturation of the resistance change could not be achieved, as several samples failed under further increases in current density.  However, the XMLD-PEEM image shown in Fig.\,4{\bf b}, captured after completing the pulse sequence depicted in panel {\bf a}, clearly reveals a N\'eel vector alignment parallel to the current direction in the central region of the sample, consistent with expectations for thermomagnetoelastic switching.
\\

{\bf Summary \& Conclusion}

In principle, all current pulse-induced magnetization switching in spintronics is associated with heating effects that generate strain. In epitaxial thin films, strain modifies the magnetocrystalline anisotropy of the investigated compound, ultimately leading to switching. Thus, when studying novel and potentially relatively weak switching mechanisms, such as SOTs acting on AFMs, it is essential to avoid confusion with the thermomagnetoelastic mechanism. 

In the context of antiferromagnetic spintronics, the thermomagnetoelastic mechanism is inherently useful, as it in principle enables the switching of all collinear AFMs with multiple easy axes. However, its speed is limited because heat and strain must propagate across the active area of the devices. Thus, achieving ultrafast processes necessitate SOT-driven dynamics.

The compound Mn$_2$Au is particularly noteworthy for its ability to exhibit both thermomagnetoelastic as well as NSOT switching within the same sample, facilitating their identification. We investigated N\'eel vector switching in two distinct thin-film geometries, enabling a clear separation of the two mechanisms. Specifically, in the 45$^{\rm o}$-cross geometry, thermally induced strain and the predicted NSOT act in opposition. For current pulse widths in the millisecond and microsecond ranges, we observed a directional dependence of the N\'eel vector reorientation that is consistent with the thermal mechanism alone. However, a transition to NSOT switching is observed for pulse widths in the nanosecond range, as required for ultrafast applications. In this regime, the N\'eel vector aligns along a specific direction perpendicular to the current, which is a characteristic feature of the NSOT and could enabling novel applications in spintronics.

Our experiments provide compelling evidence that the current-induced bulk NSOT, predicted ten years ago \cite{Zel14}, is sufficiently large to enable N{\'e}el vector switching in Mn$_2$Au devices. Therefore, despite the identification of often dominant thermomagnetoelastic switching, our results strongly support the fundamental viability of SOT-driven switching in AFM spintronics.
 
\section{Methods}

All experimental data shown in this manuscript were obtained investigating Mn$_2$Au(001)(45~nm) thin films grown epitaxially on Ta(001)(13~nm)/Mo(001)(20 ~nm) double buffer layers on MgO(100) substrates. All layers were deposited by magnetron sputtering by the process described in detail in ref.\,\cite{Bom20}.  The samples were capped with 2~nm of polycrystalline SiN$_x$ to protect them from oxidation. Optical lithography and ion beam etching were used to pattern the films into cross-shaped 50~$\Omega$ matched coplanar wave guide structures as shown in the Supplemental Material (Fig.\,S4).

Antiferromagnetic domain imaging was performed by combining photoemission electron microscopy with x-ray magnetic linear dichroism (XMLD-PEEM) at the Mn L$_{2,3}$ absorption edge at the PEEM endstations at beamline MAXPEEM at MAX IV, and beamline I06 at Diamond Light Source, and at the SIM beamline of the Swiss Light Source. The XMLD effect at the Mn L$_{2,3}$ absorption edge in Mn$_2$Au was established in previous work \cite{Sap17, Bom21}. For x-ray polarisation along a Mn$_2$Au $\langle 110 \rangle$ direction, the Mn L$_{2,3}$ XMLD spectrum shows a minimum and a maximum located at the absorption edge $E_{max}$ and at \SI{0.8}{\eV} below the edge. At MAXPEEM, the x-ray beam has normal incidence at the sample surface. At I06 and SIM, the x-ray beam is incident at a grazing angle of \SI{16}{\degree}. The XMCD-PEEM images of Ni$_{80}$Fe$_{20}$ were obtained at I06 based on the x-ray magnetic circular dichroism at the Fe absorption edge. 

In-situ electrical manipulation was performed at MAXPEEM using the pulse functions of a Keithley2601B-PULSE source for 10~$\mu$s and at SIM using a Keithley 2430 (at Diamond) sourcemeters for 1~ms pulses, integrated into the X-PEEM setup.

The ex-situ pulsing in the 100~ns to 2.5~ns range was performed using a Siglent SDG7102A 2-channel arbitrary waveform generator driving two Mini-Circuits LZY-22+ HF amplifiers. The in this frequency range required 50$\Omega$ impedance matching was obtained by using a coplanar waveguide design of the Mn$_2$Au thin films as shown in Fig.\,S4 of the Supplementary Information, by bonding the sample with short wire-bonds onto a coplanar waveguide sample holder and by using exclusively coaxial cable connections.

The ex-situ resistance measurements (Fig.\,4) were performed using a Keithley 6220 precision current source with a probe current of 100~$\mu$A and a Keithley 2182A Nanovoltmeter in Delta mode averaging over 100 measurements to obtain one data point. For automatising the pulse - probe sequence Teledyne CR33S30 HF-relais were used. 
\\

{\bf Acknowledgements}

We acknowledge funding by the Deutsche Forschungsgemeinschaft (DFG, German Research Foundation) - TRR 173 - 268565370 (project A05 (M.J.), with contributions from A01 and B02, by the EU HORIZON-CL4-2021-DIGITAL-EMERGING-01-14 programme under grant agreement No. 101070287 (SWAN-on-chip) and by the TopDyn Center (M.K.). We acknowledge MAX IV Laboratory for time on beamline MAXPEEM under Proposals 20230305 and 20240253 (M.J.), Diamond Light Source for time on beamline I06 under proposal MM37862-1 (M.J.), and Swiss Light Source for time on beamline SIM (M.J.). Research conducted at MAX IV, a Swedish national user facility, is supported by the Swedish Research council under contract 2018-07152, the Swedish Governmental Agency for Innovation Systems under contract 2018-04969, and Formas under contract 2019-02496. 
\\

{\bf Author contributions}
M.J. wrote the paper and coordinated the project; J.B., G.O.G., S.R. and M.J. prepared the samples and performed the current pulsing and resistance measurements; J.B., G.O.G., S.R., L.O., M.F. and M.J. performed the X-PEEM investigations supported by Y.R.N, E.G., F.M., and A.K.; H.S. supported the development of the HF-pulsing set-up; M.K. contributed to the discussion of the results and provided input. 
\\

{\bf Competing interests:}

 The authors declare no competing financial interests.

\newpage

\section{Supplemental Material}

{\bf I. AFM domain configurations of Mn$_2$Au(001)}

Figure S1 shows an XMLD-PEEM image illustrating a typical domain configuration in as-grown Mn$_2$Au(001) epitaxial thin films prior to the application of current pulses. After switching the same sample with a 10~$\mu$s current pulse, the resulting domain configuration is depicted in Fig.\,1{\bf g} of the main manuscript.
\begin{figure}
\includegraphics[width=0.6\columnwidth]{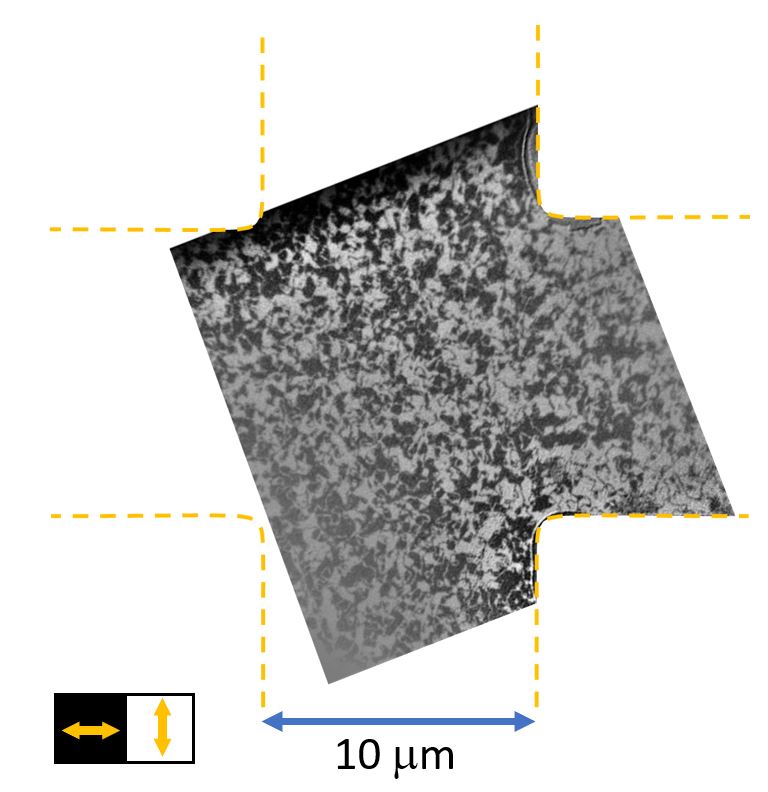}
\caption{\label{Fig1} 
S1: XMLD-PEEM image of the AFM domain pattern of a Mn$_2$Au thin film before any current injection.}
\end{figure}

The longest current pulses used to switch 45$^{\rm o}$ Mn$_2$Au crosses had a width of 1~ms. An XMLD-PEEM image of a sample switched under these conditions is shown in Fig.\,S2. This result aligns with Fig.\,1{\bf h} of the main manuscript, as in both cases, the N\'eel vector becomes aligned parallel to the current direction, indicating thermomagnetoelastic switching. The required current density for 1~ms pulses is slightly lower than that for 10~$\mu$s pulses, amounting to $2.6\times 10^{11}$~${\rm A/m^2}$.
\begin{figure}
\includegraphics[width=0.65\columnwidth]{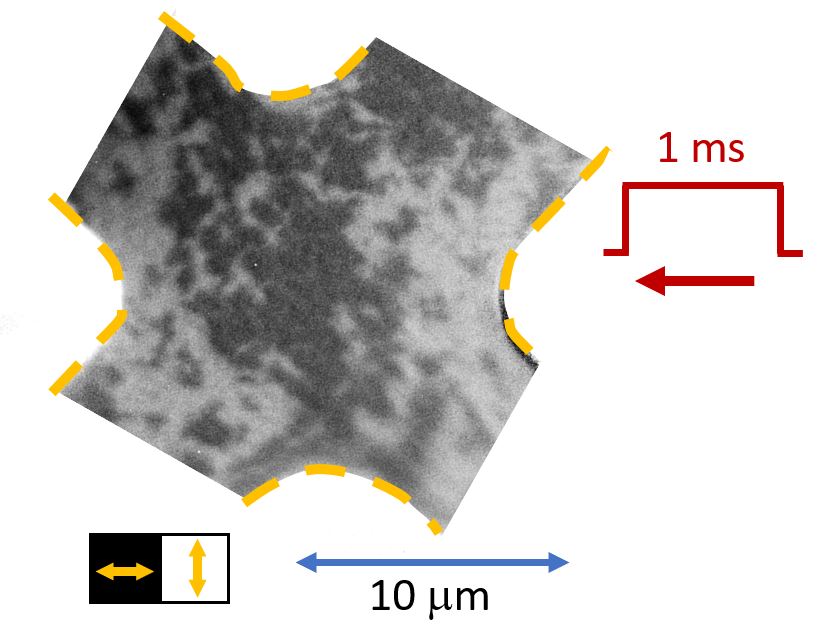}
\caption{\label{Fig1} 
S2: XMLD-PEEM image of a Mn$_2$Au 45$^{\rm o}$ cross switched with a 1~ms current pulse.}
\end{figure} 
\\

{\bf II. Full N\'eel vector orientation vs 180$^{\rm o}$ domains}

As described in the main text of the manuscript, NSOT results in an alignment of the N\'eel vector along a particular direction, thereby removing 180$^{\rm o}$-domain walls. In contrast, the thermomagnetoelestic switching mechanism generates only an axis, along which the N\'eel vector aligns both parallel as well as antiparallel, thereby generating 180$^{\rm o}$-domains.
Experimentally, the XMLD contrast mechanism is not able to distinguish between antiparallel directions of the N\'eel vector.
However, 180$^{\rm o}$-domains are separated by 180$^{\rm o}$ domain walls and these produce an XMLD contrast because the N\'eel vector rotates within the easy c-plane of Mn$_2$Au. Separating 180$^{\rm o}$ domains, such domain walls have to form closed loops. Many of those loops are visible in the center of both 45° crosses shown in Fig.\,2{\bf h} of the main manuscript. Supplementary Fig.\,S3 shows enlarged images of the centers of these from Fig.\,2 of the main manuscript. Panel {\bf a} shows the configuration obtained after a 10$\mu$s current pulse where inside and outside the 180$^{\rm o}$ closed loops is not distinguishable corresponding to the same total area with the N\'eel vectors pointing {\it right} and {\it left}. The situation is different in panel {\bf b} showing the configuration obtained after a 2.5~ns current pulse: There, the few remaining 180$^{\rm o}$ domain wall loops are much smaller in diameter and well separated. The vast majority of the central area of the device forms one connected, i.e.\,not separated by 180$^{\rm o}$-domain walls, area. This means that in this area the N\'eel vector points in the same direction, providing strong evidence for NSOT switching.
\begin{figure*}
\includegraphics[width=1.5\columnwidth]{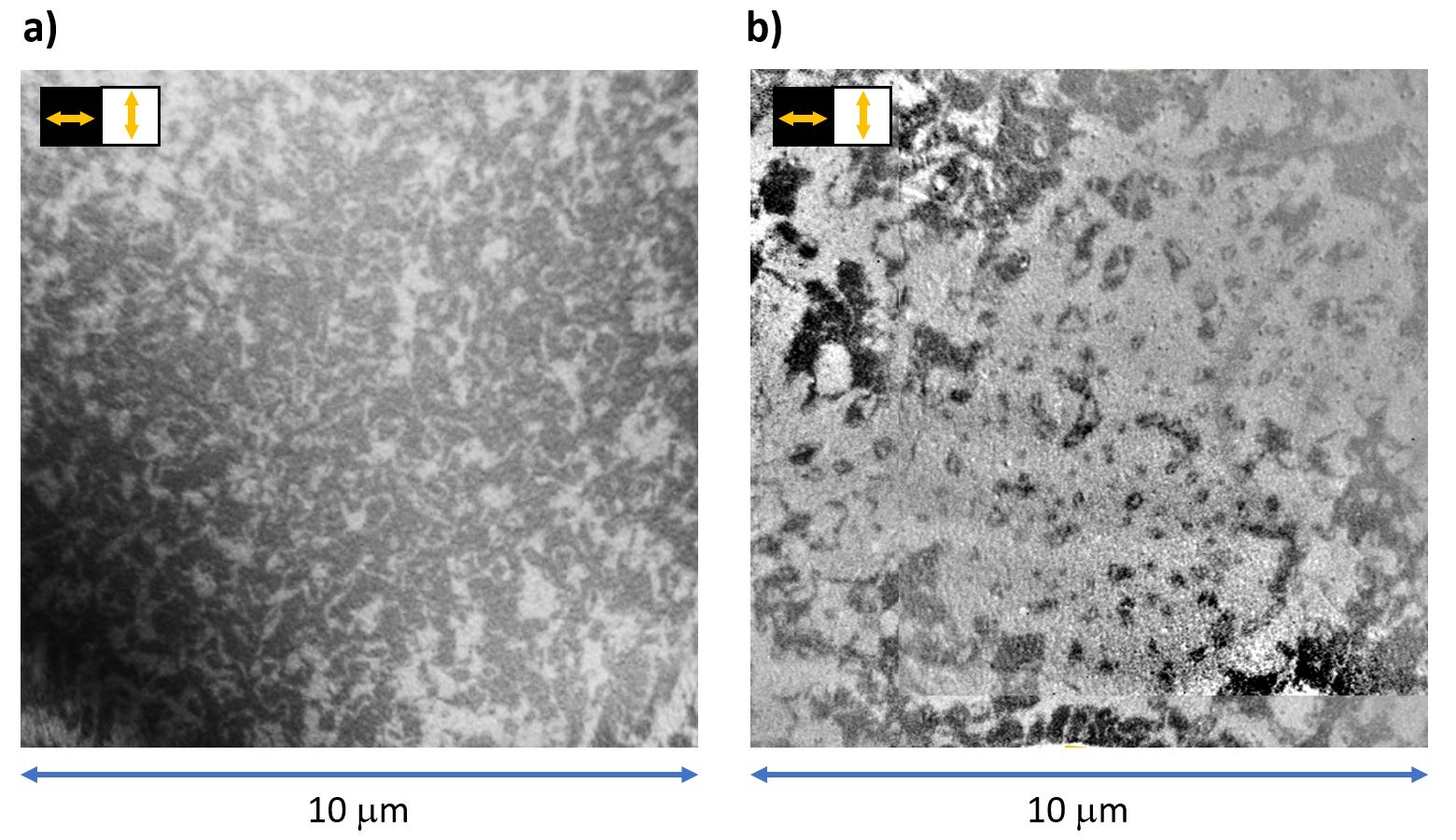}
\caption{\label{Fig1} 
S3: Enlarged XMLD-PEEM images of the centers of Mn$_2$Au 45$^{\rm o}$-crosses corresponding to Fig.\,2 of the main manuscript. Panel {\bf a}: Switched with a 10~$\mu$s current pulse. Panel {\bf b}: Switched with a 2.5~ns current pulse.}
\end{figure*} 
\\

{\bf III. Sample geometry \& pulse connections}

For current pulse durations in the nanosecond range, proper impedance matching to 50~$\Omega$) is required. Therefore, the samples were patterned using a coplanar waveguide geometry, as shown in Fig.\,S4. For pulsing with the current flowing horizontally through the center of the 45$^{\rm o}$ cross, two current pulses with positive polarity are simultaneously applied to the right arms, while two current pulses with negative polarity are applied to the left arms.

\begin{figure}
\includegraphics[width=0.7\columnwidth]{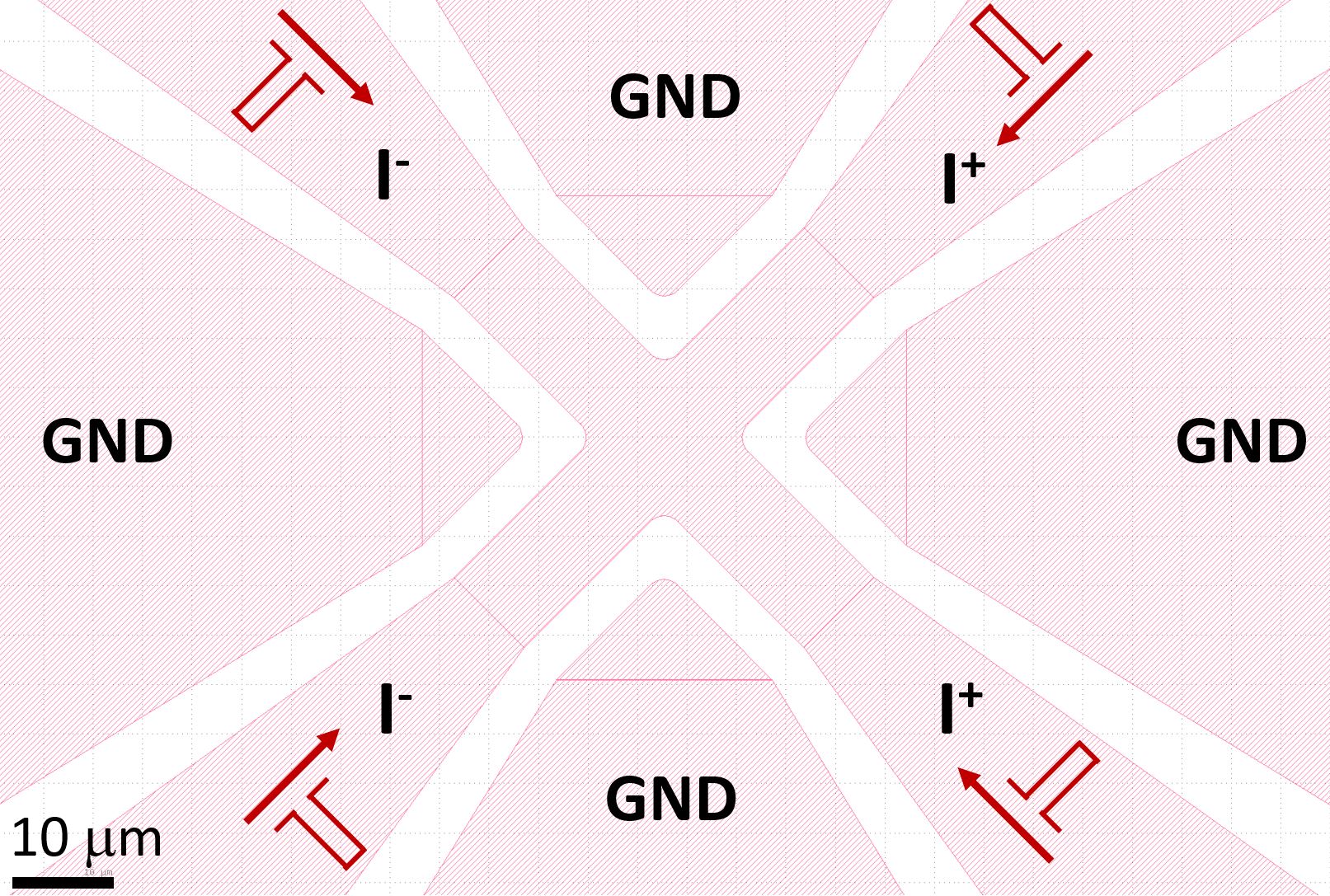}
\caption{\label{Fig1} 
S4: Sample geometry for nanosecond current pulsing with electrical connections.}
\end{figure} 

\vspace{10pt}
{\bf IV. Parameters of the COMSOL simulations of temperature and strain}

The COMSOL Multiphysics for 3D finite element modelling of temperature and strain profiles is similar to the one presented in [Zha19]. It considers the electric power generated by a current pulse, which heats the sample resulting in a thermal expansion. The thermal expansion coefficients and temperatures of the metallic thin film and of the insulating substrate are different, resulting in straining (anisotropic thermal expansion) of the Mn$_2$Au layer, which is clamped to the MgO substrate.
The parameters of the simulations are specific heat capacity $c$, thermal conductivity $k$, thermal expansion coefficient $\alpha$, Young’s modulus $E$, and Poisson ratio $\nu$. These values are published for MgO, but not available for Mn$_2$Au. Thus, we have used typical values for metals for the simulation of the Mn$_2$Au layer: $c=200$~J/(kg~K), $k=400$~W/(m~K), $\alpha=6 \times 10^{-6}~K^{-1}$, $E=170$~GPa, $\nu=0.38$. For the electrical conductivity of the Mn$_2$Au thin film, we took the measured value of $1.04 \times 10^{7}$~S/m. For the insulating MgO, the corresponding values are published: $c=900$~J/(kg~K) [Wat93], $k=48$~W/(m~K) [Wat93], $\alpha=10 \times 10^{-6}~K^{-1}$ [Rao14], $E=250$~GPa [Ezh10], $\nu=0.18$ [Ezh10]. Further parameters are the heat transfer coefficients, which have been chosen as 10~W/(m$^2$K) for the device-air interface and 500~W/(m$^2$K) for the device-sample holder interface [Zha19]. As boundary condition for the temperature distribution, we assume that the top and bottom of our sample are in contact with a 300~K environment.
\\

[Ezh10]	Ezhil, A. M., Jayachandran, M., Sanjeeviraja, C., Fabrication Techniques and Material Properties of Dielectric MgO Thin Films — A Status Review. CIRP J. Manuf. Sci. Technol. {\bf 2} 92 (2010).

[Rao14]	Rao, A. S. M., Narender, K., Studies on Thermophysical Properties of CaO and MgO by $\gamma$-Ray Attenuation. J. Thermodyn. {\bf 1}, 123478 (2014). 

[Wat93]	Watanabe, H Thermal Constants for Ni, NiO, MgO, MnO and Co0 at Low Temperatures. Thermochimica Acta, {\bf 218}, 365 (1993).

[Zha19] Zhang, P., Finley, J., Safi, T., Liu, L. Quantitative Study on Current-Induced Effect in an Antiferromagnet Insulator/Pt Bilayer Film. {\it Phys.\,Rev.\,Lett.} {\bf 123}, 247206 (2019).

\end{document}